\documentclass[12pt,preprint]{aastex}

\newcommand{\gtabout}{\ga}
\newcommand{\Hubble}{{\it Hubble Space Telescope}}
\newcommand{\HST}{{\it HST}}
\newcommand{\Ibanoglu}{\.Ibano\u{g}lu}
\newcommand{\IUE}{{\it IUE}}
\newcommand{\kms}{{\>\rm km\>s^{-1}}}
\newcommand{\ltabout}{\la}

\shortauthors{Bond et al.}
\shorttitle{CME's in V471 Tau}

\begin{document}

\title{Detection of Coronal Mass Ejections in V471 Tauri with the {\it
Hubble Space Telescope}\altaffilmark{1}}

\author{Howard E. Bond\altaffilmark{2},
D. J. Mullan\altaffilmark{3},
M. Sean O'Brien\altaffilmark{2},
and Edward M. Sion\altaffilmark{4}}

\altaffiltext{1}{Based on observations with 
the NASA/ESA {\it Hubble Space Telescope}, obtained at the Space
Telescope Science Institute, which is operated by AURA, Inc., under
NASA contract NAS5-26555.}

\altaffiltext{2}{Space Telescope Science Institute, 
		 3700 San Martin Dr.,
		 Baltimore, MD 21218; 
		 bond@stsci.edu,
		 obrien@stsci.edu} 

\altaffiltext{3}{Bartol Research Institute, 
		University of Delaware, 
		Newark, DE 19716;
	        \hbox{mullan@bxclu.bartol.udel.edu}}

\altaffiltext{4}{Department of Astronomy \& Astrophysics, 
		 Villanova University,
		 Villanova, PA 19085;
		 edward.sion@villanova.edu}

\begin{abstract}

V471~Tauri, an eclipsing system consisting of a hot DA white dwarf (WD) and a
dK2 companion in a 12.5-hour orbit, is the prototype of the pre-cataclysmic
binaries. The late-type component is magnetically active, due to its being
constrained to rotate synchronously with the short orbital period. During a
program of  ultraviolet spectroscopy of V471~Tau, carried out with the Goddard
High Resolution Spectrograph (GHRS) onboard the {\it Hubble Space Telescope},
we serendipitously  detected two episodes in which transient absorptions in the
\ion{Si}{3} 1206~$\mbox{\AA}$ resonance line appeared suddenly, on a timescale
of $\ltabout$2~min. The observations were taken in a narrow spectral region
around Lyman-$\alpha$, and were all obtained near the two quadratures of the
binary orbit, i.e., at maximum projected separation ($\sim$$3.3\,R_\sun$)  of
the WD and K~star.

We suggest that these transient features arise when coronal mass ejections
(CME's) from the K2 dwarf pass across the line of sight to the WD\null.
Estimates of the velocities, densities, and masses of the events in V471~Tau
are generally consistent with the properties of solar CME's. Given our
detection of 2 events during 6.8~hr of GHRS observing, along with a
consideration of the restricted range of latitudes and longitudes on the
K~star's surface that can give rise to trajectories passing in front of the WD
as seen from Earth, we estimate that the active V471~Tau dK star emits some
100--500 CME's per day, as compared to $\sim$1--3 per day for the Sun. The K
dwarf's mass-loss rate associated with CME's is at least
(5--$25)\times10^{-14}\,M_{\odot}\rm\, yr^{-1}$, but it may well be orders of
magnitude higher if most of the silicon is in ionization states other than
\ion{Si}{3}.

\end{abstract}


\section{Introduction}

It has long been difficult to estimate with confidence the mass-loss rates from
cool dwarf stars.  On the basis of X-ray emission, there is clear evidence that
coronae exist around many cool dwarfs. Based on analogy with the Sun, it is
likely that thermally driven winds emerge from these coronae. However, the
column densities of these winds are so small that  direct detection is usually
very difficult.

The thermally driven wind is, however, not the only means by which the Sun
loses mass. There are also episodic ejections of material into the wind. These
coronal mass ejections (CME's) contribute on average  some 16\% to the
solar-wind mass flux at solar maximum \cite{jac93}. The CME's occur when
magnetic configurations in the corona can no longer find an equilibrium with
the ambient plasma.

Because CME's are intrinsically magnetic in nature,  it is expected that in
stars where magnetic activity levels exceed those in the Sun, CME's should be
larger in scale, and occur more frequently, than in the Sun. In particular,
lower-main-sequence stars belonging to the category of flare stars are sites of
particularly high magnetic activity. The energy released in flares on such
stars may exceed by several orders of  magnitude the energy that is released in
the largest solar flares. In such magnetically active stars, we anticipate that
CME's might contribute significantly more than 16\% to the total mass flux. In
this paper, we report direct spectroscopic evidence that points strongly to the
existence of  CME's from a well-known magnetically active star. 

Our target is the late-type flare star in the V471~Tauri binary system, a
member of the Hyades star cluster. V471~Tau is an eclipsing binary with a
period of 0.52~day, consisting of a cool main-sequence star (dK2)  and a hot
white dwarf (DA1.5). Orbiting at only  $\sim$3.4 stellar radii from the dK
star, the hot white dwarf (WD) provides a superb UV  background source for
probing the environment  above the cool star's surface. In 1986, X-rays from
the WD were found to be rotationally modulated with a period of 9.25 minutes
(Jensen et al.\ 1986), and it was hypothesized that the modulation might be due
to  accretion of heavy elements onto a magnetic polar cap on the WD\null. 
Direct spectroscopic confirmation of the accretion hypothesis was obtained when
Sion et al.\ (1998) discovered a rotationally modulated, Zeeman-split UV
metallic absorption line in the photospheric spectrum of the WD. 

The material that  accretes onto the WD in V471 Tau is presumed to originate in
a wind from the dK2 companion. Reliable evidence that the K2 dwarf is
magnetically active is provided by optical spectroscopic and photometric
flares  which are ``common and intense" (Young 1976; Young, Rottler, \& 
Skumanich 1991; see also Beavers et al.\ 1979), by X-ray flares (Young et al.\
1983), and by radio flaring activity (Patterson, Caillault, \& Skillman 1993;
Nicholls \& Storey 1999).  The high levels of magnetic activity in this
particular K2 dwarf are attributed to fast rotation: the star is tidally locked
to the short orbital period of the binary. This leads to an angular velocity 
for the K2 dwarf, $\Omega_{\rm dK}$,  that is some 50 times greater than the
solar value, $\Omega_\sun$. As a result, dynamo action is expected to be much
more efficient than in the Sun.

Associated with enhanced magnetic activity, we expect an increase in the
efficiency of mass ejections in the form of CME's.  In support of this, we note
that Young (1976) has pointed out a possible connection between mass-loss
events in V471 Tau and episodes of activity on the K2 star.  Moreover,
transient absorptions in \ion{Mg}{2} in \IUE\/ spectra of V471~Tau were
reported by Mullan et al.\ (1989), and interpreted as  evidence for CME's.
Based on the \ion{Mg}{2} line strengths, Mullan et al.\ estimated the mass-loss
rate associated with the CME's to be at least $10^{-11} \, M_{\odot} \rm\,
yr^{-1}$.

Eclipse timings show that the orbital period of V471~Tau changes with time
(Tunca et al.\ 1979). If the period changes are secular in nature and are due
simply to mass loss from one component, the rate of period change may be used
to determine mean mass-loss rates.  This argument led to an estimate of
$\dot{M}_{\rm ecl}$ of order $10^{-7} \, M_\sun \rm\, yr^{-1}$ for V471~Tau
\cite{tun79}.   However, it is alternatively possible that the apparent period
changes are strictly periodic and are due to light-travel-time effects in a
triple system, as has been argued by \Ibanoglu\ et al.\ (1994) and Guinan \&
Ribas (2001).  Yet another possibility, perhaps the most likely, is that the
orbital period modulations in V471~Tau and other active binaries arise from the
magnetic activity. A specific mechanism is the  gravitational interaction
between the orbit and variations in shape of the active component in the
system, as discussed by, e.g., Applegate (1992), who argues that it is
plausible for the case of V471~Tau. In view of these alternatives, the cited
value of $\dot{M}_{\rm ecl}$ is probably a gross overestimate.  In a later
section, we will support this conclusion using an independent line of argument:
we will show that the mass-loss rate from the K2 dwarf cannot exceed  $10^{-8}
\, M_\sun \rm\, yr^{-1}$.

In the present paper,  we report on UV observations of V471~Tau obtained with
the Goddard High Resolution Spectrograph (GHRS)  onboard the \Hubble\/
(\HST\/)\null. The spectra  show clear evidence of transient absorptions in the
1206.51~$\mbox{\AA}$ resonance line of \ion{Si}{3}, which we attribute to CME's
passing in front of the WD component. The \ion{Si}{3} line is representative of
hotter material than \ion{Mg}{2}, and we therefore expect  that  the
\ion{Si}{3} line  more nearly reflects the coronal conditions in V471~Tau. To
the extent that this expectation is valid, the present study provides a more
reliable analysis of  CME properties than that based on the earlier  \IUE\/
data \cite{mul89}.

We note that V471~Tau is a detached binary, in which neither star  fills its
Roche lobe. Therefore, the mechanism that drives mass loss from the K2 dwarf
has nothing to do with Roche-lobe overflow. Instead, the mass loss  from the
dK2 star must depend for its origin on the same processes to which a single
star has access. For that reason, we may meaningfully consider processes that
are known to operate in the Sun in our attempt to interpret our observations of
V471~Tau.

\section{Observations}

The GHRS observations that form the basis of the present paper have been
described in detail by Sion et al.\ (1998) and O'Brien et al.\ (2001). The
observing was optimized primarily to determine the radial velocities of the WD
component, and also to search for spectroscopic variations on the rotational
period of the WD; our discovery of transient CME events was thus entirely
serendipitous. Briefly,  in 1993, 1994, and 1995, in the course of 12 \HST\/
visits, 128 individual spectra of V471 Tau were obtained centered on the
wavelength of Ly$\alpha$. At these wavelengths, the hot WD completely dominates
the spectrum, aside from a narrow chromospheric Ly$\alpha$ emission line from
the K~dwarf. The dispersion was 0.07~\AA~diode$^{-1}$, and the wavelength range
covered was 36~\AA, broad enough to include, by chance, the  \ion{Si}{3}
1206~$\mbox{\AA}$ resonance line. In 1994-95, there were eight \HST\/ visits,
during which each individual sub-exposure had a duration of 127~s, chosen in
order to resolve the 9.25-minute rotational period of the WD into four time
bins. During each visit, a series of 16 sub-exposures was taken, lasting
(including overheads between sub-exposures) a total of 36.3~min.  During the
four visits in 1993, the sub-exposures were 101~s, and the total observation
duration per visit was 29.0~min. Each visit was planned to occur near orbital
quadratures, i.e., at the maximum projected separation  between the WD and the
dK2 components as seen from Earth. The total time that the star was under
observation during the 12 visits  was 6.8~hours.

Transient absorptions in \ion{Si}{3} $\lambda$1206~$\mbox{\AA}$ were seen twice
during the course of the observations.  Fig.~1 shows a pictorial representation
of the spectra obtained during the more dramatic event, which occurred on 1994
Oct 20.  Sixteen individual 127-s exposures are stacked in Fig.~1 from top to
bottom in order of increasing time, in order to simulate a single-trailed
photographic spectrogram. Very broad photospheric Ly$\alpha$ absorption wings
extend across the spectrum, at constant strength in all 16 spectra, and there
are emission features from the dK chromosphere and a deep interstellar
Ly$\alpha$ absorption centered in the core of the hydrogen line.  In the top
four spectra, there is no sign of the \ion{Si}{3} line, but this line appeared
suddenly in absorption in the fifth spectrum.  The absorption was strong as
soon as it appeared, and it remained comparably strong throughout the remainder
of the 36-min series. This observation was taken at orbital phase~0.82 (WD
receding and dK2 approaching). When the next exposures at the same quadrature 
were obtained 3 days later, the \ion{Si}{3} absorption had disappeared.  As
noted above, at this orbital phase, the projected separation of the stars is at
its maximum of about 3.4~dK radii (O'Brien et al.\ 2001).

A second, similar but less pronounced transient \ion{Si}{3} event (not
illustrated here) was detected near the end of another visit on 1994 Oct 17, at
orbital phase~0.77. This transient made a sudden appearance in the 14th
sub-exposure, and remained detectable during the final three sub-exposures of
this visit. When the next visit at the same orbital phase occurred 2 days
later, the \ion{Si}{3} feature had again disappeared.

In Fig.~2, we plot the equivalent width (EW) of the \ion{Si}{3} line versus
time during the 36~minutes of observation on 1994 Oct 20. The EW was below
detectability ($\ltabout$40~m\AA) during the first four sub-exposures. Then, in
the fifth sub-exposure, the line appeared at EW
$\approx$110~m$\mbox{\AA}$\null. The line grew in strength to a maximum of
$\approx$275~m$\mbox{\AA}$ in the eighth sub-exposure, and  then fluctuated
approximately between these  limits for the remainder of the visit. 

The heliocentric radial velocity of the \ion{Si}{3} line during the 1994 Oct
20 absorption event (determined by summing the 12 spectra showing the line, and
measuring the wavelength of the line center) was $+176\pm3\kms$. At this time,
the center-of-mass velocity of the WD  was $+188\pm3\kms$, according to the
spectroscopic orbit of O'Brien et al.\ (2001).

For the event on 1994 Oct 17 the \ion{Si}{3} radial velocity was
$+220\pm10\kms$, as compared with the WD velocity of $+200\pm3\kms$. 

Thus both transient features  had radial velocities within about
$\pm$$20\kms$  of the WD radial velocity.  This indicates that, in the
reference frame which rotates with the orbital motion, the transient absorber
crossed the line of sight to the WD almost exactly transversely.

We note that the sudden, strong absorption feature we discuss here is distinct
from the much weaker and broader absorption \ion{Si}{3} line arising in the
photospheric surface spot on the WD discovered by Sion et al.\ (1998): the
velocities are significantly different (note that a photospheric line would
have an additional gravitational redshift of $+49.9\kms$), a surface ``patch"
of \ion{Si}{3} ions would be rotationally modulated on the 9.25-min WD rotation
period, and a photospheric line would show Zeeman splitting (see Sion et~al.).

\section{Interpretation of the Transient Events as CME's}

We will now discuss the transient absorption events in terms of CME's passing
in front of the WD.

\subsection{Transverse Velocity of the Absorbing Material}

The facts that both of the transient features appeared suddenly,  were not
present in the immediately preceding spectra, and then persisted for the rest
of the observation, imply that the absorbing material must have covered the
entire disk of the WD within a time interval $\Delta t$ of less than the
exposure time of 127~s.



With a WD radius, $R_{\rm WD}$, of $7440\pm490$~km (O'Brien et al.\ 2001), 
this requires that the absorbing material must have traversed our line of sight
to the WD at a projected speed, $V_{\rm proj}$, of at least $2R_{\rm WD}/\Delta
t \simeq 120 \kms$.  Given the suddenness with which the feature appeared, the
true velocity of the absorber, $V_{\rm abs}$, almost certainly exceeded
$120\kms$, perhaps by a considerable factor.

\subsection{Lower Limit on the Linear Extent of the Absorbing Structure}

The transient absorption on 1994 Oct 20 persisted until the end of the 36-min
observing sequence (see Figs.~1 and~2). Thus the duration of the absorption
event, $\Delta t_{\rm abs}$, was at least 1600~s.  Given that the absorbing
feature moved across the line of sight to the WD with a speed of $V_{\rm abs}$,
we conclude that the extent of the absorber  in projection, $\Delta L$, was at
least $ V_{\rm abs} \, \Delta t_{\rm abs}$, i.e., $\Delta L \ge
1.9\times10^5$~km. Thus, the  absorbing feature had a transverse dimension,
$L_{\rm CME}$, that  exceeded  $\sim$$0.28\,R_\sun$.


The dK2 star in V471~Tau has a radius  that is slightly less than that of the
Sun: $R_{\rm dK} \simeq 0.96 \, R_\sun$ \cite{obr01}. Thus, the above lower
limit on $L_{\rm CME}$ in terms of the dK star radius is of order $0.29\,R_{\rm
dK}$.

%
%
%

\subsection{Probability of Detecting an Absorber in Transit}

Only those CME's that emerge from the K dwarf in V471~Tau and pass in front of
the WD will create absorption features in its spectrum. 

Moreover, the two transient features that we did detect during the 6.8~hours of
observing were seen to move nearly transversely across the WD line of sight in
the orbital reference frame.

The question we need to address is this: under what launch conditions will a
CME from the K dwarf satisfy the above conditions?

To answer this question, we must consider the orbital mechanics of an object 
that moves in the gravitational field of a rotating binary with mass parameter
$\mu = m_2/(m_1+m_2)$, where conventionally $m_1$ is taken to refer to the more
massive component. Thus for the case of V471~Tau we take $m_1$ to refer to the
mass of the dK star and $m_2$ to refer to the mass of the WD\null. O'Brien et
al.\ (2001) have used the same GHRS observations described here to determine
the spectroscopic orbit of the WD component, and have derived masses of
$m_1=0.93\,M_\sun$ and $m_2=0.84\,M_\sun$. We thus  have $\mu = 0.475$.

Using the notation of Gould (1957) for the restricted three-body problem,  we
adopt a three-dimensional $(x,y,z)$ frame of reference rotating with the orbit,
with the $x$-axis chosen to lie along the line of centers and directed from the
dK primary to the WD secondary (see Fig.~3).  The sense of rotation is denoted
by the arrows close to the $\Omega$ symbol in Fig.~3.  With the origin at the
center of gravity,  the primary and secondary stars are fixed at $x_1 = -\mu$
and $x_2 = (1-\mu)$, respectively.  For simplicity, we restrict our attention
to motions in the orbital $(x,y)$ plane. 

The equations that we integrate are  
$$\ddot{x} = 2\dot{y} + x - (1-\mu)(x-x_1)/r_1^3 - \mu(x-x_2)/r_2^3 \, ,$$ 
$$\ddot{y} = -2\dot{x} + y - (1-\mu)y/r_1^3 - \mu y/r_2^3 $$ 
\cite{gou57}, where $r_{1,2}^2 = y^2 + (x-x_{1,2})^2$.  In these equations, 
the    unit of distance is the separation, $a$, between the stars ($a \simeq
3.30 \, R_\sun \simeq 3.43 \, R_{\rm dK}$), and the unit of time is such that
$P_{\rm orb}=2\pi$. The unit of velocity thus becomes $2\pi a/P_{\rm orb}$, or
about $320\kms$. Following Gould, we checked the numerical accuracy of each
integration by evaluating Jacobi's constant, and found it to be invariant to
within one part in $10^9$ after 100 time units.

Each integration starts by launching a ``test particle" radially outward from
the surface of the K2 dwarf (with radius 0.29 in the normalized units used
here). The launch point is chosen to lie on the equator, at a longitude
$\theta$ relative to the line of centers. Because of the complicated
gravitational potential, the trajectory is very sensitive to both $\theta$ and
$V_0$, the initial launch velocity. For examples of the great variety of
trajectories that can occur, see Gould (1959), especially her Fig.~8 in which
$\mu$ has the value 0.5, close to that of V471~Tau. If we knew how to make
allowance for magnetic forces, we would probably find trajectories of even
greater complexity. However, in the present study, we restrict attention to the
mechanics that are  controlled by gravitation and rotation alone, which we can
characterize  with high precision in V471~Tau where the stellar and orbital
parameters are well known.\footnote{With increasing distance from the K star,
magnetic forces on the CME will weaken rapidly (since $B^2 \propto r^{-6}$),
whereas centrifugal forces increase in strength ($\propto r$). As a result, the
CME in effect becomes magnetically decoupled from the K star at a relatively
short distance from the star. Beyond that point, the trajectory is expected to
be dominated by gravitation and rotation. For our present purposes, we assume
that the decoupling of the CME occurs so close to the K star that, apart from
controlling the launch velocity, magnetic fields have no significant effects on
the subsequent motion.}

We plot in Fig.~3 a series of trajectories for ejections with a range of $V_0$
values and three different launch points on the dK surface. Shown as solid
lines are launches from the sub-WD point (longitude $\theta = 0$), with $V_0$
ranging from 1.2 to 2.0 (i.e., $\sim$380--$640\kms$ for the parameters of
V471~Tau). As noted, the plot is in the rotating frame of reference. The line
of sight (LOS) to the Earth at the quadrature where the WD is receding from us
is shown. The trajectories launched from the sub-WD point all pass close to the
WD, and are favorable for creating absorption in the WD spectrum at this
quadrature. Moreover, in all of these cases  the velocity of the particle is
almost entirely in the $x$-direction, i.e., transverse to our LOS\null.
(Specifically, all of the solid lines plotted in Fig.~3 have $V_y$ of less than
$\pm$0.07 units of velocity during the WD occultation. For the parameters of
V471 Tau, this corresponds to less than $\pm$$22\kms$.) This family of 
trajectories thus agrees quite well with the velocity properties actually
observed for our two transient events. Moreover, such trajectories would not be
able to produce absorption events at orbital phase 0.25, again in agreement
with the fact that we did not observe any transients at that phase.

It should also be noted that these trajectories all leave the system after
passing in front of the WD, i.e., they represent mass that is lost from the
K~star and from the binary.

Several optical observers, including Young, Skumanich, \& Paylor (1988) and
Bois, Lanning, \& Mochnacki (1991), have reported that H$\alpha$ emission tends
to be strong when V471~Tau is near orbital phase 0.5 (i.e., when the sub-WD
point faces the Earth).   This might suggest that there is enhanced activity
near the sub-WD point, which would favor launches from near this region on the
star. On the other hand, the strong emission is probably related as well to the
UV irradiation of the facing hemisphere by the hot WD\null. In any case, as
noted by Young, Rottler, \& Skumanich (1991), the H$\alpha$ profiles are highly
complex and variable, even at the same orbital phase, so that variable surface
activity is clearly present over and above the constant effect of irradiation.

It is possible to find other launch points and velocities that also
lead to trajectories passing in front of the WD at phase 0.75. Fig.~3,  for
example, shows (as dotted lines) a family of launches from $\theta=+45^\circ$,
with $V_0=1.4$,  1.8, and 2.0, several of which do pass in front of the WD as
desired. However, launches from $\theta=-45^\circ$, shown as dot-dash lines in
Fig.~3, do not pass in front of the WD as seen from Earth at phase 0.75.

In order to evaluate the probability of CME detections such as those we
observed, we searched in $\theta$ and $V_0$ to determine those launch
conditions that give rise to occultations  in which $V_y$ is no larger than
$\pm$0.07. For purposes of occultation, we allow the ejecta to  have a finite
extent  corresponding to the lower limit mentioned above ($\sim$$0.3\,R_{\rm
dK}$):  as a result, occultation extends from slightly before to slightly
after  the instant at which the center of the ejecta transits the WD\null. We
demand that $V_y$ have values in the range  $\pm$0.07 at some point during this
extended occultation.

Results of this search are shown in Fig.~4, in which combinations of $\theta$
and $V_0$ that satisfy these criteria are plotted as $Y$'s. For sufficiently
low $V_0$, the ejecta do not escape from the K dwarf at all;  they simply fall
back onto its surface. In the V471~Tau system, this happens at $V_{0,\rm min} =
1.06$, shown by a vertical line in Fig.~4.  For $V_0$ in excess of $V_{0,\rm
min}$,  we find that launches satisfy our criteria over a certain range of
$\theta$ values.  The limits of this range are outlined roughly by dashed
lines: outside the dashed lines, the launches are not relevant in the present
context. The allowed launch range $\Delta\theta$ is broadest (some
50--$70^\circ$ wide) for $V_0 \approx 1.4$--1.6.  For very fast $V_0$, the
ejecta move past the WD at an increasingly large angle, such that it becomes
difficult to satisfy the $V_y$ criterion. As a result, only a narrow launch
window (some 8--$15^\circ$ wide) is possible for $V_0 \gtabout 2$.

It is true that, in the rotating frame, the paths of most CME's eventually 
wrap around the binary: therefore many trajectories may in principle eventually
intercept the LOS to the WD\null. However, this generally happens at such a
great  distance from the WD that the chances of occultation (as seen by us) are
much smaller than in the closest approaches shown in Figs.~3-4, and also the
gas density may well have fallen significantly at that time.

If CME's with the optimal velocities $V_0 = 1.4$--1.6 (i.e.,
$\sim$450--$500\kms$) were ejected uniformly at all longitudes around the
equator of the K2 dwarf, the probability that we would detect absorption in the
WD near phase 0.75 would be very roughly $\Delta \theta/360^\circ \approx
0.15$--0.2. Thus, for every CME that is in principle detectable, we will miss
$N_{\rm long}\approx 5$--7 others because they are launched from the ``wrong"
longitude. $N_{\rm long}$  will be even larger than this if $V_0$ is outside
the optimal range.

We will also miss CME's that are ejected out of the orbital plane. In our
computer modelling, we considered only ejections from zero latitude. Because of
the finite extent of $L_{\rm CME}$, however,  an ejection at non-zero latitude
$\lambda$ could still occult the WD if $L_{\rm CME}/a \leq \sin\lambda$. With
$L_{\rm CME} \gtabout 0.3 R_{\rm dK}$, and $a \simeq 3.4\,R_{\rm dK}$,
$\lambda$ exceeds $\sim$$5^\circ$.

If $L_{\rm CME}$ were  $\approx$$R_{\rm dK}$,  $\lambda$ would be of order
$17^\circ$. Thus, if the CME's are also launched from random latitudes, for
every CME that is detectable, we may miss $N_{\rm lat} \approx 3.2$--11 
because they are launched from the ``wrong" latitude. These are upper limits on
$N_{\rm lat}$, since Coriolis forces eventually force even high-latitude
ejections to approach the equatorial plane, so that occultation may eventually
be possible. As a result, a range of crudely 3--10 is not implausible for
$N_{\rm lat}$ in the  V471~Tau system.


Thus, assuming randomly distributed launch points,  for every ejection that we
``catch" in absorption in V471~Tau at phase 0.75,  we ``miss" at least $N_{\rm
miss} = N_{\rm long} \, N_{\rm lat} \approx 15$--70 CME's.

Of course, it is difficult to guarantee that  CME's emerge from the K dwarf in
a spherically symmetric manner. If the emergence of CME's is non-spherically
symmetric,  the above estimates of $N_{\rm miss}$ will be modified. In the case
of the Sun, where activity is confined to low latitudes, most CME's are
observed to be launched  within $40^\circ$ of the equator \cite{hil77}. If this
tendency were to occur in V471~Tau,  the above estimates of $N_{\rm miss}$
should be reduced by factors of about 1.5. However, it is not obvious that the
latitudinal distribution of activity on a  rapidly rotating K star is the same
as on the Sun: for example, such a star may  differ from the Sun in having
polar spots and in having differential rotation of opposite sign from that in
the Sun (e.g., Vogt \& Hatzes 1991).  On a star with polar spots, the CME's
might even be ejected preferentially in the polar direction: in such a case,
$N_{\rm miss}$ would be even larger than we estimated above. Indeed, Doppler
mapping of the dK star in V471~Tau by Ramseyer, Hatzes, \& Jablonski (1995)
does show spots on the surface of the star at high latitudes.

\subsection{Rate of CME ejection}

Two transients were detected in V471 Tau in the course of 6.8 hours of our GHRS
observations.  This corresponds to a mean observable transient rate of
$\approx$7 per day.  Correcting for the geometric factors ($N_{\rm miss}$) in 
the previous subsection, and assuming that the transient rate that we observed
is typical of the time-averaged rate, the total rate of CME emergence from the
K star is, very crudely, of order 100--500 per day. 

\subsection{Comparison with Solar CME Properties}

We note that when CME's emerge from the Sun, the outward velocities have lower
limits of about $100 \kms$ \cite {hil77}. This is close to the lower limit we
have estimated on $V_{\rm abs}$ in V471~Tau.

However, the mean velocities of solar CME's at various epochs are considerably
higher, 350--$470\kms$ \cite{hun97}. In this regard, we note that the optimal
launch velocities of CME's in V471~Tau (as far as being able to create with
high probability absorptions in the WD spectrum similar to those we observed)
was found to be $\sim$1.4--1.6 velocity units, or about 450--$500\kms$ (cf.\
Fig.~4). This overlaps with the high end of the solar CME range of velocities.

As regards the sizes of CME's in the Sun, we note that the mean  latitudinal
extents of solar CME's  at various epochs are found to be 42--$47^\circ$
\cite{hun97}.  Very few events are larger in angular extent than $65^\circ$
\cite{hil77}. Thus solar CME's near their source in most cases have linear
dimensions that are comparable to $1\,R_\sun$, with very few having larger
linear dimensions. If a scaling by stellar radius applies to  CME's from the K
star in V471~Tau, then $L_{\rm CME}$ should not exceed about $R_{\rm dK}$ in
linear extent.  This limit on CME dimensions does not disagree with the
observational constraints discussed above.

We note that inhomogeneity in solar CME's is relevant to the V471~Tau case. The
internal structure of CME's is far from uniform. For example, two CME's
reported by Wood et al.\  (1999) exhibited highly striated interiors, with many
different strands of material apparently intersecting one another in
complicated patterns. Each strand had a width of at least $10^4$~km, and was
presumably associated with a different flux rope in the original magnetic
configuration. If strands of similar widths are present in CME's from the K2
star in V471~Tau, then each individual strand could cover the WD (with
diameter  less than $10^4$~km) as a separate event. The presence of multiple
strands could explain why the \ion{Si}{3} absorption line fluctuates
significantly with time once the event has begun, as shown in Fig.~2.

In summary, it seems that in terms of speed, dimension, and internal
complexity,  the transient absorbing features in  V471~Tau are not inconsistent
with the properties of solar CME's. Below, we shall see that in terms of
overall mass, the feature  also resembles a solar CME\null. However, for the
Sun, even near solar maximum, the rate of CME's does not exceed about 3 per
day: near solar minimum, the number is closer to 1 per day \cite{hil77}. The
rate of CME ejections from the active V471~Tau K dwarf evidently exceeds the
solar value by a considerable factor, roughly 2 orders of magnitude.

\subsection{Another Measure of the Activity Level on V471 Tau}

Another measure of the level of magnetic activity on a star  involves the size
and/or number of cool spots on the surface. In the Sun, as the number of spots
waxes and wanes in the course of the 11-year cycle, the largest spots and
faculae have areas of order $10^{-3}$ times the visible disk area. As a result
of non-exact cancellation between spot darkenings and facular brightenings, the
solar ``constant" varies during a solar cycle with an amplitude of about
0.05\%, or 0.0005~mag \cite{fou90}.

In contrast to these small variations, the brightness of the K dwarf in
V471~Tau  varies by  as much as 0.2~mag over timescales of tens of years
(\Ibanoglu\ 1978; Tunca et al.\ 1979; \Ibanoglu\ et al.\ 1994). If these
variations can be ascribed to waxing and waning of spots/faculae analogous to
the solar activity cycle, then the areal dimensions of spots/faculae on the K
dwarf in V471 Tau must exceed those in the Sun by factors of 200-400. In fact,
radiometric calculations presented by O'Brien et al.\ (2001), along with direct
Doppler imaging by Ramseyer et al.\ (1995), confirm that about 25\% of the
surface area of the star is covered by starspots, some 250 times that of the
Sun.


\subsection{Dynamo Activity in V471 Tau}

We note that our estimates of the excess of the CME rates above solar values
(100--500) and our estimates of the excess of spottedness above solar values
($\sim$250) are reasonably similar.

Is such a high level of activity possible in the context of dynamo activity? To
answer this, we recall that the angular velocity $\Omega_{\rm dK}$ of the K2
dwarf in V471~Tau is some 50 times larger than the  angular velocity of the
Sun. Presumably the level of magnetic activity is determined by the maximum
magnetic field strength on the surface of the star. Now, in a turbulent 
dynamo, the amplitude of the maximum field  depends on the mechanism that gives
rise to non-linear limiting. With one particular choice of limiting mechanism,
the maximum field strength is predicted to scale as $\Omega^{3/2}$
\cite{kip73}. This prediction suggests that magnetic fields in the K2 dwarf in
V471~Tau may exceed the solar values by a factor of $\sim$350.

We suggest that the excesses in CME rate and in spotted activity in V471~Tau
are related to this  dynamo-related excess in maximum field strength.


What effects might arise from these strong fields? Presumably the coronal
heating process will be highly efficient. In the Sun, enhanced magnetic field
strengths, $B$, are correlated with enhanced coronal densities, $n$. Because of
this correlation, the Alfven speed $V_A \propto B/n^{1/2}$  does not
necessarily vary greatly from one region of the corona to another. Now, when a
CME is created through loss of magnetic equilibrium, the velocity of ejection
may be related to the local Alfven speed. In such a scenario, the speeds of
CME's  may exhibit a well-defined mean in spite of large variations of $B$
value from one part of the corona to another. Analogously, the presence of
strong fields in V471~Tau does not exclude the possibility that $V_A$ is
comparable to solar $V_A$\null. This may explain why the optimal ejection speed
in V471 Tau overlaps with the mean range of solar CME speeds.

\section{Estimating the Mass-Loss Rate}

In this section, we use the properties of the transient absorption features to 
set limits on the CME mass-loss rate from the K2 dwarf.

\subsection{Column Depth of the Absorbing Feature}


We wish to estimate $N_{\rm Si}$,  the number of silicon nuclei per cm$^2$
along the line of sight during a transient absorption. To be specific, we
consider the equivalent width of the transient  absorption in the \ion{Si}{3}
1206~$\mbox{\AA}$ line at the time of maximum absorption: $W_A \approx
0.3$~$\mbox{\AA}$ (see Fig.~2).  To convert this to a column density, we assume
that conditions are optically thin; this seems to be a reasonable approximation
in view of the relative weakness of the transient absorption. On the linear
portion of the curve of growth, $W_A$ is related to  the column density of
ground-state \ion{Si}{3} ions, denoted $N_{\rm Si\,III}$,  as follows 
\cite{spi68}: $$W_A = 8.85 \times 10^{-13} \, \lambda_{\mu}^2 \, f \, N_{\rm
Si\,III} \, .$$ Here, $W_A$ is in \AA,  $\lambda_{\mu} = 0.1206$ is the
wavelength in microns, and $f$ is the oscillator strength of the
1206~$\mbox{\AA}$ line.  According to Allen (1963), $f = 1.9$. Inserting $W_A =
0.3$~\AA, we find $N_{\rm Si\,III} = 1.2 \times 10^{13}\,\rm cm^{-2}$. The
total column density of silicon nuclei, $N_{\rm Si}$, exceeds $N_{\rm Si\,III}$
by a factor $1/\psi(\rm Si\,III)$, where $\psi(\rm Si\,III)$ is the fractional
abundance of silicon in the  \ion{Si}{3} ground state. The numerical value of
$\psi(\rm Si\,III)$  is related to the temperature of the source region from
which the CME material emerged.

Assuming a cosmic silicon abundance  of  $\rm H/Si \approx 3 \times 10^4$ by
number \cite{all63}, we find that the column density of hydrogen in the
transient absorption is $N_{\rm H} \approx 4 \times 10^{17} /\psi(\rm Si\,III)
\rm\, cm^{-2}$. 

\subsection{Number Density in the Absorbing Material}

Denoting the linear extent of the absorbing material $L_{\rm CME}$, we recall
from \S3.2 that its value is at least  $\sim$$0.3\,R_\sun \simeq
2\times10^{10}\rm\,cm$. The volumetric number density of hydrogens, $n_{\rm
H}$, in the absorbing material is thus given by $n_{\rm H} \approx 2 \times
10^7 \, (0.3 \,R_\sun/L_{\rm CME}) /\psi(\rm Si\,III) \rm\,cm^{-3}$.

We shall find below (see \S4.5) that it is possible to set a lower limit of
$\sim$$10^{-5}$ on the value of $\psi(\rm Si\,III)$. If $L_{\rm CME}\approx
1R_\sun$, we can therefore obtain an approximate upper limit on the number
density of the absorbing material of $n_{\rm H} \approx 6 \times
10^{11}\rm\,cm^{-3}$.

Is this limit reasonable for a detached binary system? To answer that,  we note
that in semi-detached binaries (where, due to Roche-lobe overflow, denser gas
streams are expected),  gas densities are found to be in the range
$10^{12}$--$10^{15}\rm\,cm^{-3}$ \cite{bat70}. These are, as expected, larger
than the upper limit we obtain here for the detached V471~Tau system.



\subsection{Mass of the Transient Absorber}

With $n_{\rm H}$ protons per cm$^3$, and linear extent of $L_{\rm CME}$, the
total mass of the 1994 Oct 20 transient absorber, $M_{\rm CME}$,  was of order
$n_{\rm H} \, m_{\rm H} \, L_{\rm CME}^3$, where $m_{\rm H}$ is the proton
mass.  Using the above estimates, we find $M_{\rm CME} \simeq 3 \times
10^{14}\,  (L_{\rm CME}/0.3\,R_\sun)^2 /\psi(\rm Si\,III) \,\, g$. If we again
assume  $L_{\rm CME} \approx 1\,R_\sun$ (analogous to the largest solar CME's),
and consider that $\psi(\rm Si\,III) \le 1$, then the CME mass for the event
seen in V471~Tau must have exceeded $3\times10^{15}$~g.

Is this mass reasonable for a CME\null? To answer this, we note that in the
case of the solar corona,  the mass distribution of CME's has an exponential
distribution with typical scale $10^{16}$~g \cite{jac93}. Thus, once again, it
is noteworthy that a  CME parameter in the Sun has a comparable value in
V471~Tau. The limiting mass of a CME from a stellar corona  is presumably
determined by the ability of the plasma and magnetic field of the corona to
store energy only up to some limiting value \cite{mul00}. In V471~Tau, the
levels of magnetic activity are larger than solar, but it is less clear what
the plasma properties might be. Thus, it is not clear whether or not the
atmosphere in V471~Tau is in principle capable of storing larger energies than
in the solar case. 

\subsection{CME Mass-Loss Rate in V471 Tau}

Combining our estimates of the numbers of ejections per day ($\sim$100--500)
from the K dwarf with $M_{\rm CME}$ from the preceding subsection, we find the
CME mass-loss rate in V471~Tau to be of order  $\dot{M}_{\rm CME} \simeq
(3$--$15) \times 10^{11} \, (L_{\rm CME}/0.3\,R_\sun)^2/\psi(\rm Si\,III)
\rm\,\, g\,s^{-1}.$

How does this compare to the rate at which the Sun loses mass in the form of
CME's? According to Jackson \& Howard (1993), the maximally active Sun ejects
CME mass at a rate of $2.7 \times 10^{11}\rm\,g \, s^{-1}$.  Thus the CME
mass-loss rate in V471~Tau is comparable to or somewhat larger than  that in
the Sun if most of the Si is in the form of \ion{Si}{3}, or could greatly
exceed the solar rate if $\psi(\rm Si\,III)$ is low.

We thus see that, although all of the input numbers are uncertain, the
principal unknown factor in determining the actual CME mass-loss rate in
V471~Tau  is the numerical value of  $\psi(\rm Si\,III)$.  We now turn to a
discussion of  this value.

\subsection{Fractional Ionization of Silicon}

The value of $\psi(\rm Si\,III)$ depends on the electron temperature, $T_e$, at
the place of CME origin. In an expanding wind, ionization states are ``frozen
in"  within a short radial distance of the stellar surface. ``Frozen-in"
conditions set in at the radial distance  where the expansion time of the wind
becomes shorter than the ionization time-scale \cite{owo83}. Beyond that radial
distance, even if the gas cools as the CME expands during its outward motion,
the ionization states will not come into local equilibrium. 

We refer to ionization equilibria calculations under chromospheric\slash
coronal conditions (e.g., Arnaud \& Rothenflug 1985, their Table~IV) to
determine the fractional abundance of \ion{Si}{3} at various temperatures. (At
our level of approximation, we ignore the small fraction, generally
$\ltabout$25\%, of \ion{Si}{3} ions that are in excited states.) If the
material in the transient absorber  started in the corona of the K star, where
electron temperatures are at least $10^6$~K, then $\psi(\rm Si\,III)$ would be
miniscule: Arnaud \&  Rothenflug's (1985) table does not extend beyond $\log
T_e = 5.6$ for \ion{Si}{3}, but if we  perform an (admittedly gross)
extrapolation from the last three entries in their table, we find that at $T_e
= 10^6$~K, $\psi(\rm Si\,III)$  may be of order $10^{-10}$. If the material
started out in the low transition region between corona and chromosphere, at
(say) 140,000~K, then  $\psi(\rm Si\,III) \approx 10^{-3}$. The highest
abundance of \ion{Si}{3} occurs at electron temperatures of $\sim$30,000~K,
where $\psi(\rm Si\,III) \approx 1$. Finally, if the material started off in
the chromosphere (at $\sim$10,000~K) then $\psi(\rm Si\,III) \approx 10^{-4}$. 

For $\psi(\rm Si\,III)\simeq1$, and $L_{\rm CME}\simeq1\,R_\sun$, the CME
mass-loss rate from the results of the previous subsection would be
(3--$15)\times10^{12}\rm\,g\,s^{-1}$ or
(5--$25)\times10^{-14}\,M_\sun\rm\,yr^{-1}$. To get fairly reasonable agreement
with the estimated lower limit of $>$$10^{-11}\,M_\sun\rm\,yr^{-1}$ from \IUE\/
observations of \ion{Mg}{2} absorption \cite{mul89}, we would need  $\psi(\rm
Si\,III)\simeq0.005$--0.025. These limits on $\psi(\rm Si\,III)$ correspond to
an origin either in the transition region (at temperatures above
$\sim$100,000~K), or conceivably in the chromosphere (at temperatures below
$\sim$14,000~K).

On the other hand, if the CME material in V471 Tau emerged from a truly coronal
structure with a very low $\psi(\rm Si\,III)$, then the mean mass-loss rate in
CME's could be several orders of magnitude larger, possibly even approaching
the early dynamical estimates of mass-loss rates as high as
$10^{-7}\,M_\sun\rm\,yr^{-1}$ based on changes in the orbital period (see \S1
above). However, as mentioned above, those dynamical estimates failed to take
into account the possibility of a third body in the system, or of modulations
of the orbital period by the magnetic activity itself.

Here we would like to point out another argument, based on accretion, which
suggests that a mass loss rate of order $10^{-7}\,M_\sun\rm\,yr^{-1}$ is
probably too large for V471~Tau. The argument proceeds as follows. The rate at
which material is being accreted onto the surface of the WD has been found to
be less than $10^{-17}\,M_{\odot}\rm\, yr^{-1}$ \cite{sio98}. As argued by
Mullan et al.\ (1991) and by Sion et al.\ (1998), this accretion rate is
several orders of magnitude smaller than what should be occurring if  purely
hydrodynamical processes were in control of the flow of the K2 stellar wind
around the WD\null.  This suggests that a mechanism of some kind (most probably
a ``magnetic propeller") is providing a highly efficient shield to protect the
surface of the WD from the passing wind. The properties of such a propeller
allow us to set a limit on the mass-loss rate.

To determine this limit, we refer to the discussion of Mullan et al.\ (1991).
If the mass-loss rate from the K2 dwarf  were to be as large  as
$10^{-7}\,M_{\odot}\rm\, yr^{-1}$, the ram pressure of this wind would be so
large that the ``magnetospheric radius," $R_m$,  would be pushed in almost to
the surface of the WD\null. Specifically, inserting a mass-loss rate of
$10^{-7}\,M_{\odot}\rm\, yr^{-1}$ into  eqs.\ (13)--(18) of Mullan et al., we
find  that $R_m$ would have a numerical value of only $5 \times 10^9 \,
B_6^{1/3}$~cm, where $B_6$ is the polar magnetic field strength in MG\null.
Since the field strength is now known to be of order 0.35~MG from Zeeman
splitting of the photospheric \ion{Si}{3} lines \cite{sio98}, $R_m$ does not
exceed $4 \times 10^9$~cm. (The numerical value is quite insensitive to $B_6$
because of the small  exponent: even if $B_6$ were to uncertain by a factor of
10, $R_m$ would be uncertain only by a factor of $\sim$2.)  The above estimate
of $R_m$ is to be compared with  the co-rotation radius,  $R_c = 9 \times
10^9$~cm (see eq.~(20a) in Mullan et al.\ 1991). (The value of $R_c$ depends
only on the WD mass and its  rotation period, both of which are reliably
known.)  We see that, with the above choice of mass-loss rate, $R_m$ is {\it
smaller\/} than $R_c$. The key point can now be appreciated: in a situation
where $R_m < R_c$, the propeller mechanism cannot operate. 

Since, in fact, the propeller appears to be operating with high efficiency
(Sion et al.\ 1998), we conclude that a mass-loss rate as large as $10^{-7} \,
M_{\odot}\rm\, yr^{-1}$ from the K dwarf can be ruled out. In order to preserve
propeller operation in the presence of a field of 0.35~MG, we find that the
mass-loss rate from the K dwarf must not exceed about $10^{-8} \,
M_{\odot}\rm\, yr^{-1}$. This provides us with a  lower limit on $\psi(\rm
Si\,III)$ of $\sim$(0.5--$2.5)\times 10^{-5}$. Using Arnaud \& Rothenflug's
(1985) tables, we find that the above lower limits on $\psi(\rm Si\,III)$ 
constrain the electron temperature in the place of CME origin to be no more
than $\sim$230,000-280,000~K.

Finally, we note that the estimates of mass-loss rate which we present  here
refer only to the component of mass loss associated with the transient events.
It is possible that the K2 dwarf also loses mass from its corona in the form of
a steady thermally driven wind. But the present data have insufficient
sensitivity to allow us to study the properties of that component directly. The
upper limit on $\dot{M}$ that we derived above from  propeller arguments refers
to the total mass loss rate, i.e.,  a combination of steady wind plus CME's.

\section{Conclusions}

The V471~Tau system (dK2+DA1.5) provides an extremely favorable opportunity to
detect  discrete episodes of mass loss from a magnetically active star, because
of the presence of a strongly UV-emitting, nearly point-like close companion
that shines through the stellar wind. Magnetic activity in the K2 dwarf is kept
at an unusually  high level because the angular rotation is tidally locked to 
the orbital value, some 50 times solar. By extension from the case of the solar
corona, we expect that the morphology of the corona in the K2 dwarf in V471~Tau
is dominated by magnetic structures.

In this paper, we report the serendipitous discovery of two transient
absorption features  of a \ion{Si}{3} resonance line, seen in \HST/GHRS
spectra. We interpret the transient absorption features in terms of material
ejected from the chromosphere or corona of the K2 star and traversing our line
of sight to the white dwarf. Our analysis indicates that the absorption
features have properties that are comparable to those of  CME's in the Sun,
including size, velocity, complex structure, and (lower limits on) mass. The
one property that shows striking differences from solar CME's is the rate of
occurrence: we argue that the K2 dwarf in V471 Tau ejects CME's some 100--500
times more frequently than does the Sun.

We show that the K dwarf's mass-loss rate associated with CME's is at least
(5--$25)\times10^{-14}\,M_{\odot}\rm\, yr^{-1}$, but it may well be
considerably higher if most of the silicon in the absorbing material is in
ionization states other than \ion{Si}{3}. 

In fact, the principal uncertainty in our analysis is the lack of reliable
knowledge of the fraction of silicon, $\psi(\rm Si\,III)$, that is in the form
of \ion{Si}{3} in the CME material; this quantity depends on the unknown
electron temperature at the point of origin of the material.  However, since an
efficient magnetic propeller appears to be at work in V471~Tau, we can set an
upper limit on the mass loss rate from the K2 dwarf of
$\sim$$10^{-8}\,M_{\odot}\rm\, yr^{-1}$. 

We have recently obtained additional UV spectroscopy of V471~Tau with the Space
Telescope Imaging Spectrograph, covering a much wider wavelength range than the
GHRS observations reported here, and these data should provide much additional
information on the excitation conditions in the wind and its mass-loss rate.

\acknowledgments HEB, MSO, and EMS acknowledge support from STScI through grant
GO-5468. DJM is supported in part by the NASA Delaware Space Grant program.


\clearpage

\begin{figure}
\begin{center}
\includegraphics[angle=-90, width=\hsize]{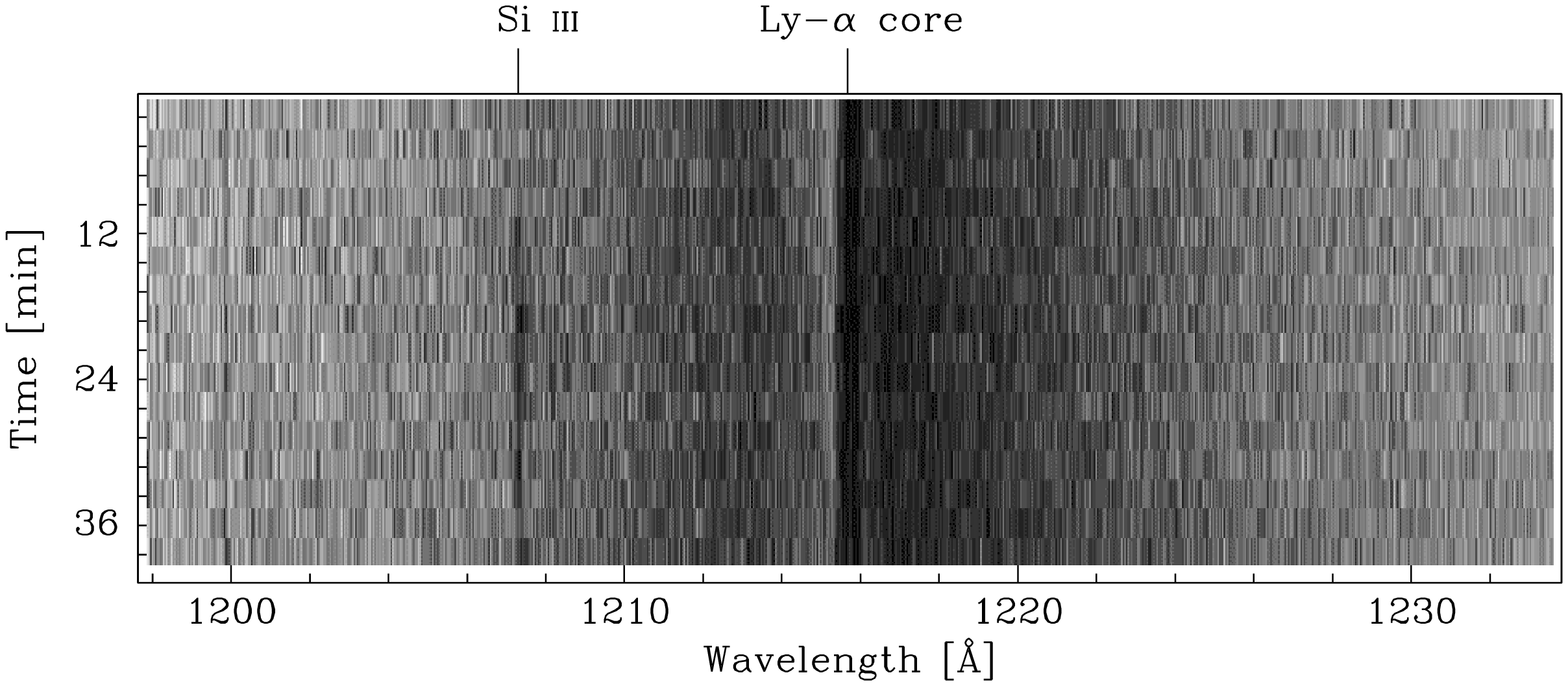}
\end{center}
\caption{Pictorial representation of the time-resolved GHRS spectra of
V471~Tau obtained on 1994 Oct 20, showing evidence for a coronal mass ejection
from the cool component of the binary projected in front of the hot white dwarf
companion. A portion of the UV spectrum centered on Ly$\alpha$ was covered at a
scale of $0.07\,\rm\mbox{\AA}\,diode^{-1}$. Sixteen spectra, each one having an
exposure time of 127~s, were taken in sequence, covering a 36-min interval. The
\ion{Si}{3} $\lambda$1206 line suddenly appeared in absorption in the fifth
subexposure and remained present for the remainder of the observation. Note
also the broad Ly$\alpha$ absorption wings from the photosphere of the white
dwarf, a blue-shifted Ly$\alpha$ emission component from the K~dwarf, and a
strong interstellar Ly$\alpha$ absorption core.\label{fig1}}
\end{figure}
\clearpage

\begin{figure}
\begin{center}
\includegraphics[angle=-90, width=\hsize]{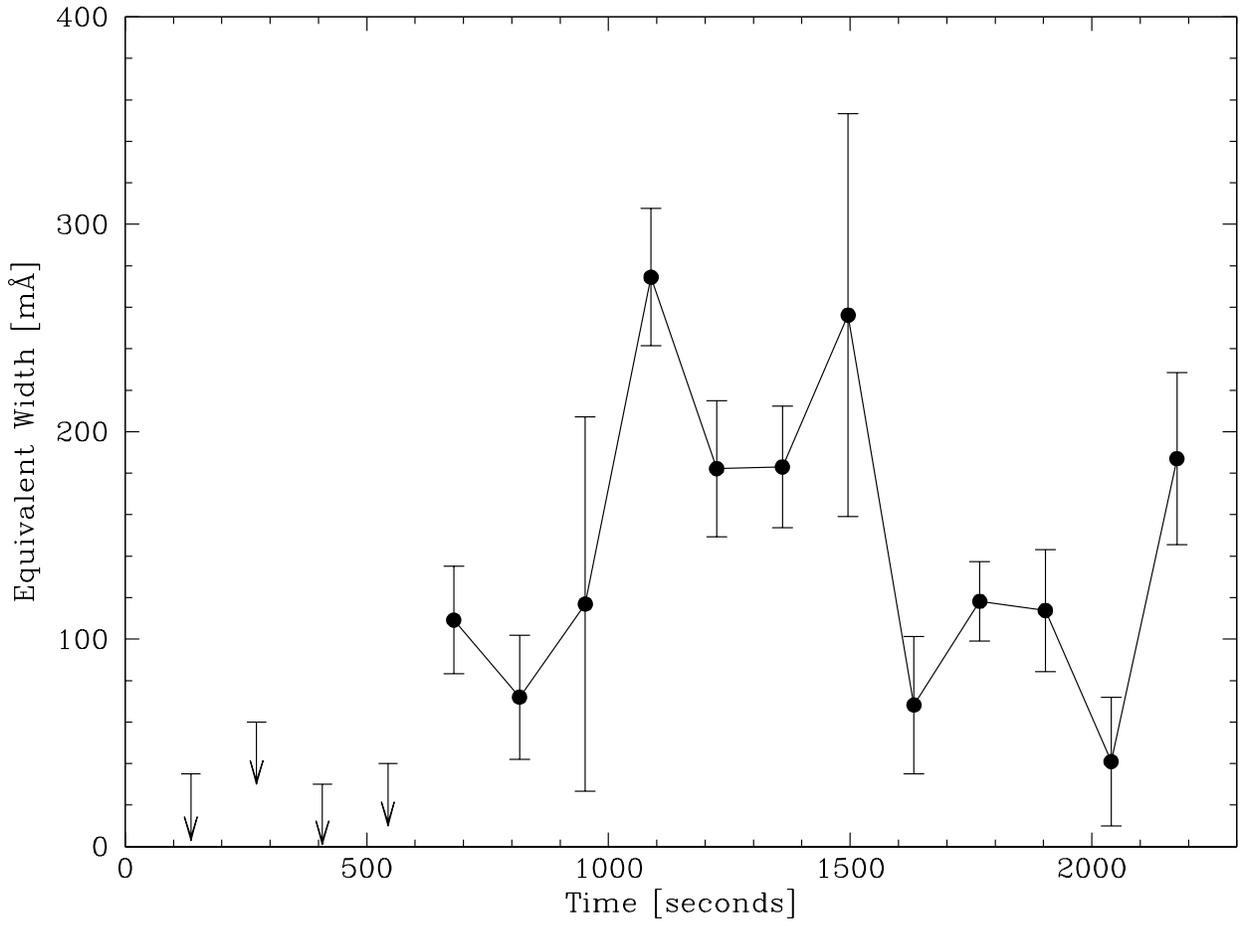}
\end{center}
\caption{Equivalent width vs.\ time for the transient \ion {Si}{3}
1206~$\mbox{\AA}$ absorption feature on 1994 Oct 20.\label{fig2}}
\end{figure}
\clearpage

\begin{figure}
\begin{center}
\includegraphics[angle=90, width=\hsize]{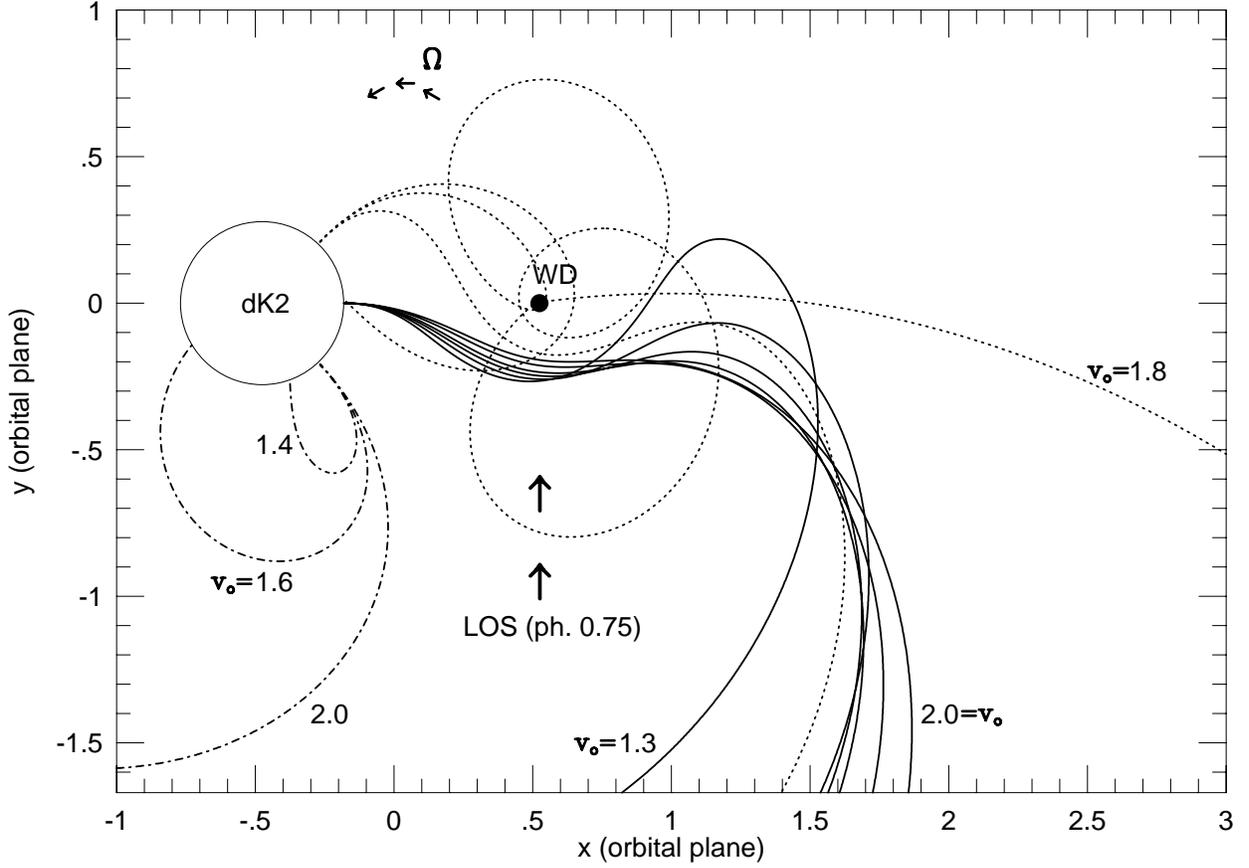}
\end{center}
\caption{Trajectories of point masses ejected from the  surface of the dK2 star
in V471 Tau (fractional mass ratio $\mu$ = 0.4757).  Only motions in the
orbital $(x,y)$ plane were considered.  The unit of distance is the orbital
separation, $a \simeq 3.432 \, R_{\rm dK}$,  between dK2 star and the white
dwarf (WD)\null. {\it Solid lines:} trajectories of masses launched from the
sub-WD point (longitude $\theta = 0$), with launch velocities of $V_0 = 1.3$,
1.4, 1.5, 1.6, 1.8, and 2.0 (in units of $320\kms$). {\it Dotted lines:}
trajectories of masses launched from $\theta= +45^\circ$ with $V_0$ = 1.4,
1.8,  and 2.0. {\it Dot-dash lines:} same for $\theta = -45^\circ$ with $V_0$ =
1.4, 1.6,  and 2.0. The figure shows that launches from near the sub-WD point
pass in front of the WD, as seen from Earth at orbital phase 0.75, with a small
$y$-velocity, as seen in the two observed transient events.\label{fig3}}
\end{figure}
\clearpage

\begin{figure}
\begin{center}
\includegraphics[angle=90, width=\hsize]{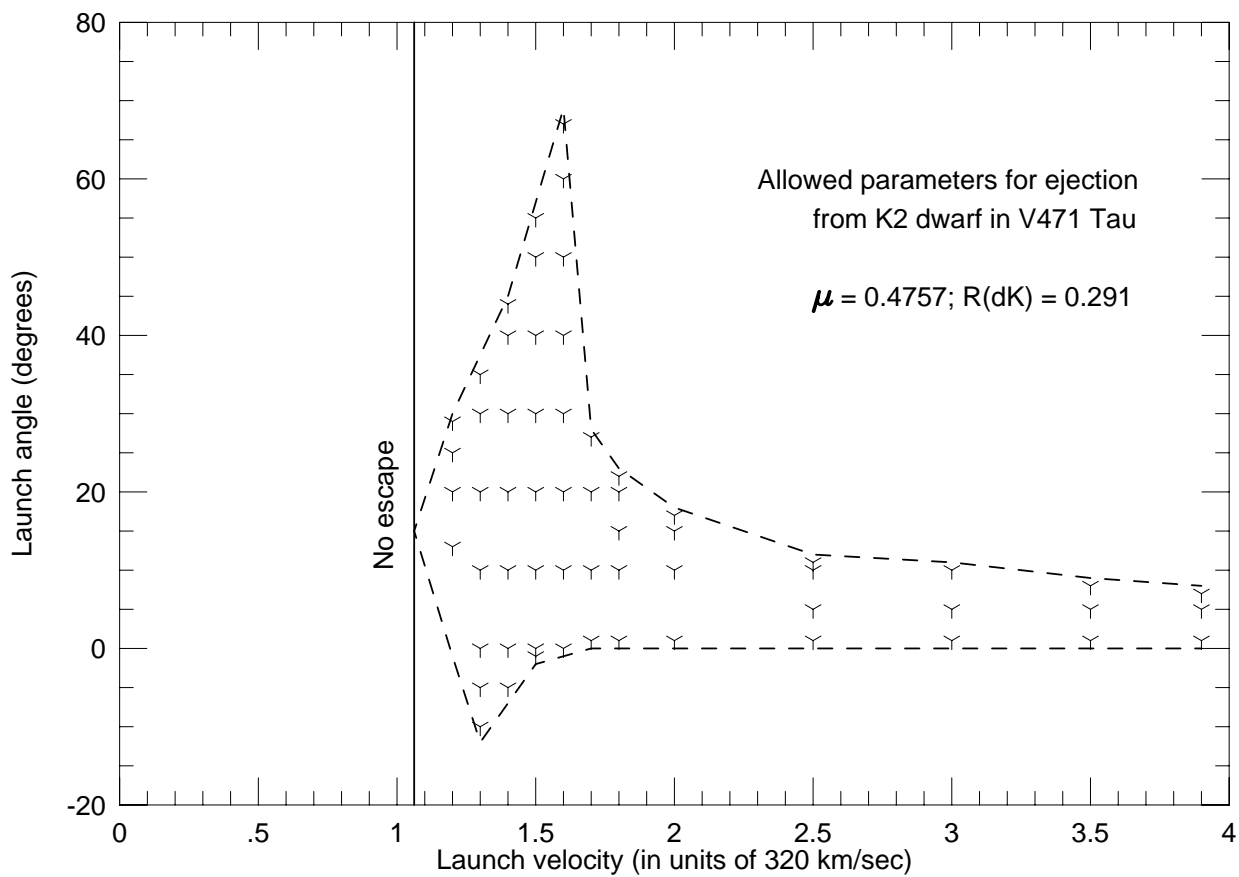}
\end{center}
\caption{Range of launch angles as a function of initial velocity, such that
when the point mass passes in front of the white dwarf (as seen by us at phase
0.75), the radial component of the point mass velocity relative to the white
dwarf is within the limits $\pm$$25\kms$ (corresponding to the properties of
the two observed absorption transients). $Y$ symbols mark trajectories that
satisfy this criterion. Launches at $V_0 \le 1.06$ (or $\sim$$340\kms$) fall
back onto the dK star, and are marked with a vertical line labelled ``no
escape.''\label{fig4}} 
\end{figure}
\clearpage

\end{document}